\documentclass{article}

\usepackage{ICLR,times}
\iclrfinalcopy
\usepackage[utf8]{inputenc} 
\usepackage[T1]{fontenc}    
\usepackage{hyperref}       
\usepackage{url}            
\usepackage{booktabs}       
\usepackage{amsfonts}       
\usepackage{nicefrac}       
\usepackage{microtype}      
\usepackage{authblk}
\usepackage[inkscapelatex=false]{svg}
\usepackage[export]{adjustbox}
\usepackage{amsmath}
\usepackage{algorithm}
\usepackage{algorithmic}
\usepackage{enumitem} 
\usepackage{subcaption}
\title{\hspace{1.2cm}
  \raisebox{-0.15\height}{\includegraphics[height=0.8cm]{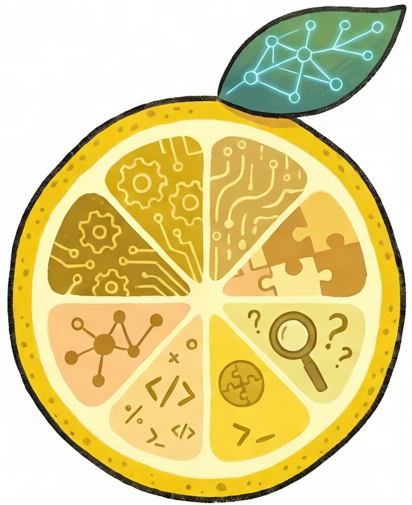}}\hspace{0.3cm}\textbf{Lemon Agent Technical Report}
}

\author{\textbf{Haipeng Jiang}\textsuperscript{\rm 1}, \textbf{Kailong Ren}\textsuperscript{\rm 1}, \textbf{Zimo Yin}\textsuperscript{\rm 1}, \textbf{Zhetao Sun}\textsuperscript{\rm 1}, \textbf{Xin Gan}\textsuperscript{\rm 1}, \textbf{Guangyi Lv}\textsuperscript{\rm 1}, \textbf{Ming He}\textsuperscript{\rm 1, $\dagger$}, \textbf{Peng Wang}\textsuperscript{\rm 1 ,$\dagger$}, \textbf{Congli Yin}\textsuperscript{\rm 1}, \textbf{Hong Pan}\textsuperscript{\rm 1}, \textbf{Changwen Zhang}\textsuperscript{\rm 1}, \textbf{Shan Tong}\textsuperscript{\rm 1}, \textbf{Zhengyu Xu}\textsuperscript{\rm 1},  \textbf{Zeping Chen}\textsuperscript{\rm 1}, \textbf{Yubin Huangfu}\textsuperscript{\rm 1}, \textbf{Yanzhi Xu}\textsuperscript{\rm 1}, \textbf{Xing Su}\textsuperscript{\rm 1}, \textbf{Qin Feng}\textsuperscript{\rm 1}, \textbf{Dong An}\textsuperscript{\rm 1}, \textbf{Jianping Fan}\textsuperscript{\rm 1}\\
\textsuperscript{\rm 1} AI Lab @ Lenovo CTO Org\\
}

\begin{document}

\renewcommand{\thefootnote}{\fnsymbol{footnote}}
\maketitle

\footnotetext[2]{Corresponding Author. Email: heming01@foxmail.com, wangpeng31@lenovo.com.}

\begin{abstract}
Recent advanced LLM-powered agent systems have exhibited their remarkable capabilities in tackling complex, long-horizon tasks. Nevertheless, they still suffer from inherent limitations in resource efficiency, context management, and multimodal perception. Based on these observations, \textbf{Lemon Agent} is introduced, a multi-agent orchestrator-worker system built on a newly proposed AgentCortex framework, which formalizes the classic Planner-Executor-Memory paradigm through an adaptive task execution mechanism. Our system integrates a hierarchical self-adaptive scheduling mechanism that operates at both the overall orchestrator layer and workers layer. This mechanism can dynamically adjust computational intensity based on task complexity. It enables orchestrator to allocate one or more workers for parallel subtask execution, while workers can further improve operational efficiency by invoking tools concurrently. By virtue of this two-tier architecture, the system achieves synergistic balance between global task coordination and local task execution, thereby optimizing resource utilization and task processing efficiency in complex scenarios. To reduce context redundancy and increase information density during parallel steps, we adopt a three-tier progressive context management strategy. To make fuller use of historical information, we propose a self-evolving memory system, which can extract multi-dimensional valid information from all historical experiences to assist in completing similar tasks. Furthermore, we provide an enhanced MCP toolset. Empirical evaluations on authoritative benchmarks demonstrate that our Lemon Agent can achieve a state-of-the-art 91.36\% overall accuracy on GAIA and secures the top position on the xbench-DeepSearch leaderboard with a score of 77+. Our experiemntal results have highlighted how the synergy between adaptive resource coordination and continuous experiential learning effectively bridges the gap between theoretical agent potential and the rigorous requirements of real world deployment.

\vspace{3pt}
Project: \url{https://github.com/Open-Lemon Agent/Lemon Agent}
\end{abstract}




\section{Introduction}
Modern AI agents have demonstrated remarkable capabilities across various tasks and application domains \citep{toolorchestra, zhang2025cosight}, where they can master complex reasoning and adaptive decision-making in highly unstructured environments \citep{2025mirothinker, zhang2025agentorchestrahierarchicalmultiagentframework, yu2025aworldorchestratingtrainingrecipe}. These breakthroughs have positioned AI autonomous agents as foundational technology for next-generation applications across critical sectors including precision healthcare, scientific discovery, and sustainable infrastructure management. Autonomous AI agents can not only deliver unprecedented efficiency gains for enterprise operations, but also democratize our daily accesses to advanced AI capabilities for real-world problem-solving \citep{DBLP:journals/corr/abs-2504-07139, JoyAgent-JDGenie}.

Even big progresses have been achieved for autonomous AI agents, they still face at least three key issues: (1) Static resource allocation across the tasks with varying complexities, where high-performance large models are overdeployed even for the trivial tasks, resulting in prohibitive operational costs \citep{DBLP:journals/corr/abs-2503-12434, DBLP:journals/corr/abs-2503-21460}. This undermines environmental adaptability in resource-constrained real-world scenarios. (2) Ineffective memory evolution strategies that depend on binary success/failure task outcomes for knowledge updates, which may fail to capture valuable insights from partial successes or format-biased failures in complex real-world tasks lacking clear ground truth \citep{DBLP:journals/corr/abs-2508-07407,DBLP:journals/corr/abs-2507-21046}. This impairs autonomous evolution through limited learning from early experiences (i.e., task execution trajectories). (3) Perceptual inadequacy when handling high-dimensional multimodal inputs. This constrains the richness and naturalness of human-agent interactions in real-world applications \citep{DBLP:journals/corr/abs-2403-12273}.

Based on these observation, we introduce \textbf{Lemon Agent}, a multi-agent system built upon AgentCortex framework to facilitate efficient and robust task execution. Lemon Agent seamlessly integrates five core structural innovations: (1) \textit{AgentCortex framework}, which implements a Planner Executor Memory paradigm to ensure industrial grade reliability and academic generality while supporting service oriented deployment; (2) \textit{a hierarchical self-adaptive scheduling mechanism}, which operates at both macro orchestration and micro execution levels to dynamically modulate computational intensity and tool parallelization according to task complexity; (3) \textit{a three-tier progressive context management strategy}, which utilizes intra tool truncation plus adaptive summarization and retroactive compression to mitigate information decay during long horizon trajectories; (4) \textit{the Self-Evolving Semantic Memory (SES Memory) module}, which enables continuous capability refinement by extracting transferable skill snippets from all execution traces regardless of final task outcomes; and (5) \textit{an enhanced MCP toolset}, which enhances the environmental interaction through specialized modules including an adaptive image understranding tool, a street view agent, a multi source search tool, a powerful file read tool, etc. These components collectively allow Lemon Agent to achieve superior performance by synthesizing adaptive resource management with continuous knowledge evolution and high fidelity environmental perception.

Experimental results have demonstrated that Lemon Agent achieves superior performance across diverse benchmarks. As of February 6, 2026, our system attains an overall accuracy of 91.36\% on the GAIA benchmark and secures the top position on the xbench-DeepSearch benchmark with an accuracy of 77+ \citep{DBLP:conf/iclr/MialonF0LS24, xbench}. These results validate our ability to manage massive search spaces and solve complex multi-step reasoning problems. Beyond academic metrics, AgentCortex framework also demonstrates its industrial-grade reliability through its deployment in the Lenovo Super Intelligent Agent. This commercial implementation achieves a transaction volume in the hundreds of millions and earns recognition as a 2025 CCF Enterprise Digitalization Outstanding Case.

Our key contributions are summarized as follows:
\begin{itemize}[left=0pt, labelsep=0.5em, itemindent=0pt, listparindent=0pt] 

    \item We introduce Lemon Agent as a unified multi-agent orchestration system built on the AgentCortex infrastructure that standardizes the Planner-Executor-Memory paradigm. Our system synthesizes a hierarchical self-adaptive scheduling mechanism and a three tier context management strategy with the SES Memory protocol for continuous skill refinement alongside an augmented tool suite featuring adaptive visual understanding, intelligent search, and precision geospatial navigation.
    
    \item We validate the exceptional performance and architectural generality of our system through extensive empirical studies on authoritative benchmarks including GAIA and xbench DeepSearch where Lemon Agent achieves state of the art results. To encourage community research and facilitate transparent validation, we provide the complete implementation as an open source project to support the advancement of autonomous agent technologies.
        
\end{itemize}

\section{Lemon Agent}
\begin{figure}[htbp]
\centering
\includegraphics[width=1.0\textwidth]{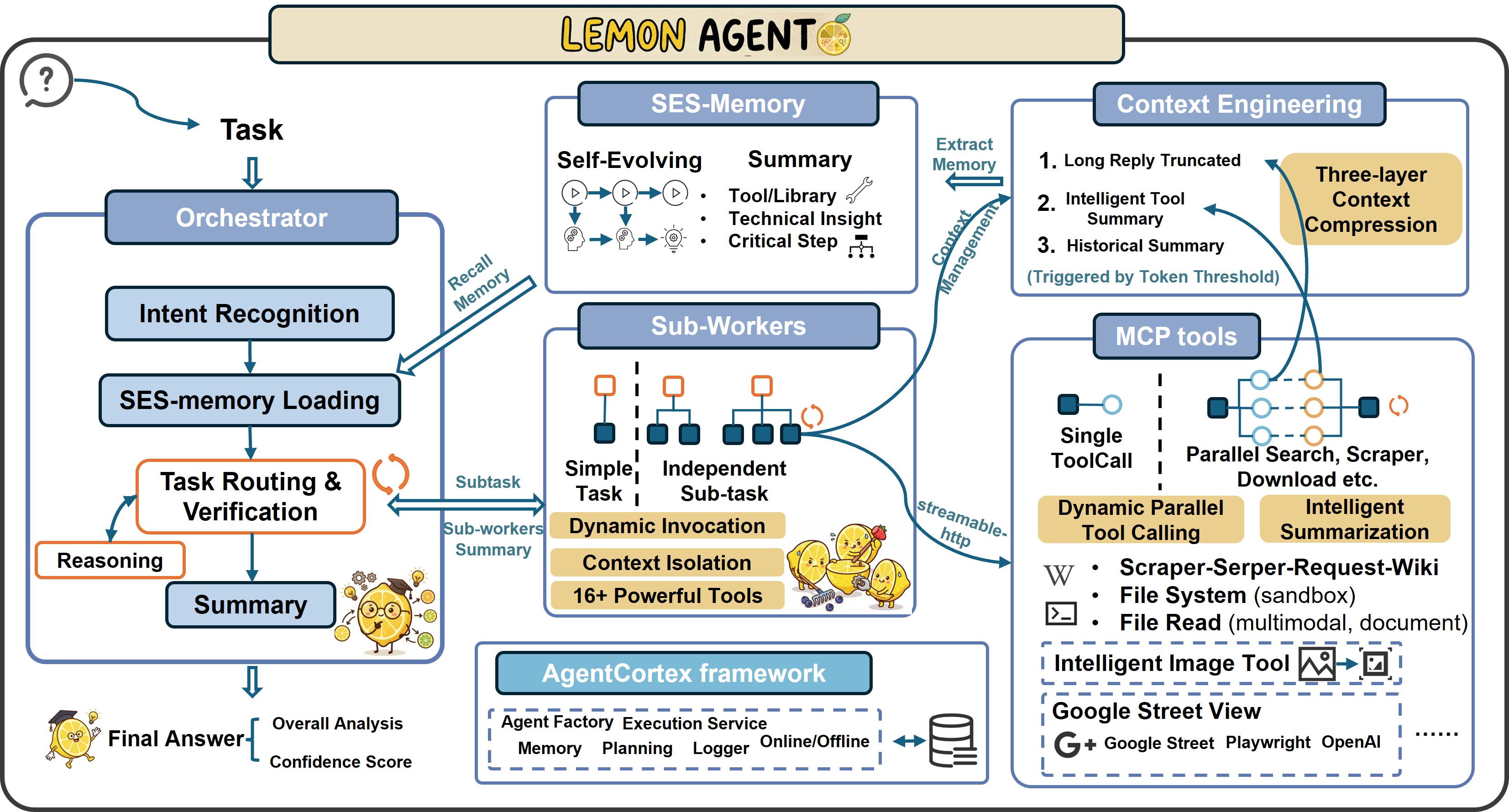}
\caption{Overview of the Lemon Agent system. Our orchestration system contains the integration of multi-agent collaboration, sub-worker cluster, self-evolving memory, and tool modules.}
\label{fig:framework}
\end{figure}

We introduce Lemon Agent, a modular multi-agent system designed for robust and efficient task execution within complex and dynamic environments. Figure~\ref{fig:framework} shows a comprehensive overview. Our system coordinates specialized sub-worker clusters through a hierarchical self-adaptive scheduling mechanism that modulates computational intensity across macro-orchestration and micro-execution levels. To maintain reasoning coherence during extended trajectories, we incorporate a three-tier progressive context management strategy and a self-evolving semantic memory system for continuous experiential learning from diverse execution traces. This integrated approach is further empowered by an augmented tool suite including adaptive visual refinement, multi-source search, and precision geospatial navigation. The details will be introduced in the following subsections.

\subsection{AgentCortex Framework}
We implement our system using AgentCortex, an agent-oriented technology framework and design paradigm designed to bridge the gap between academic research and real-world product deployment. This infrastructure provides a robust and modular foundation for building autonomous systems by offering the following technical features:

\begin{itemize}[left=0pt, labelsep=0.5em, itemindent=0pt, listparindent=0pt]
  \item \textbf{Modular Architecture.} The framework decomposes an intelligent agent into well-defined components, including intent understanding, task decomposition, planning, tool execution, knowledge retrieval, memory management, and task summarization. These components are connected through unified and highly abstracted interfaces, enabling low coupling, high flexibility, and sustainable system evolution.

  \item \textbf{Technical Convergence.} A key strength of AgentCortex is its ability to harmonize algorithmic innovation with engineering rigor. Researchers can rapidly integrate state-of-the-art or self-developed algorithms into a complete agent pipeline for experimentation, while maintaining a direct path to production-ready implementation.

  \item \textbf{Production-Grade Infrastructure.} The framework provides built-in, enterprise-level capabilities, including a high-performance microservice engine, comprehensive logging and monitoring infrastructure, and robust database access mechanisms. This ensures that agent systems can be deployed reliably and at scale in diverse real-world scenarios.
\end{itemize}

The practical effectiveness of AgentCortex lies in its ability to significantly shorten the path from experimental prototypes to production-ready intelligent agents. By aligning research workflows with industrial engineering practices, the framework allows for the seamless transition of complex agent behaviors into stable commercial operations. These achievements validate the advanced architectural design and industrial-grade reliability of AgentCortex across intricate and diverse task environments.

\subsection{Hierarchical Self-Adaptive Scheduling}
To optimize resource allocation and execution efficiency, Lemon Agent employs a hierarchical self-adaptive scheduling mechanism that operates at both the macro-orchestration and micro-execution levels. This dual-layer adaptivity allows the system to dynamically modulate its computational intensity based on the inherent complexity and structural requirements of the task. By balancing the breadth of parallel exploration with the precision of sequential reasoning, the framework achieves high performance across diverse workloads while maintaining a sustainable resource footprint. The scheduling logic is partitioned into the following two distinct dimensions of adaptivity:

\subsubsection{Macro-Level Dynamic Orchestrator-workers}
The main agent determines the necessity of multi-expert collaboration based on the structural independence of the global goal. For straightforward tasks, the system maintains a parsimonious configuration by preferring a single sub-worker invocation to minimize inter-agent communication overhead and inference latency. However, if a task is identified as possessing multiple orthogonal facets or independent sub-goals, the main agent activates a collaborative ensemble of specialized sub-workers. This macro-level modulation ensures that the system only scales its expert resources when the complexity of the task justifies the additional computational expenditure.

\subsubsection{Micro-Level Dynamic Tool Parallelization}
Within each sub-worker, the system implements a granular execution strategy that dynamically adjusts the number of parallel tool calls between one and five instances. The degree of parallelization is determined by the specific functional requirements of the sub-task. In information-intensive scenarios such as news retrieval, the sub-worker may execute multiple concurrent search engine queries followed by parallelized web crawling to accelerate data acquisition. Similarly, for multimodal tasks involving image recognition, the agent triggers simultaneous downloads and concurrent visual analysis of multiple images to ensure rapid evidence synthesis. Conversely, the scheduling mechanism prioritizes a sequential, single-tool execution mode when encountering complex reasoning chains or tasks with high inter-step dependencies. This setting for singular invocation in logical-heavy scenarios can preserve the integrity of the reasoning trajectory.

\subsection{Hierarchical Context Management Strategy}
Lemon Agent employs a three-tier progressive context compression mechanism. This strategy is designed to mitigate the risks of information decay and context window overflow, which are common in multi-step trajectories involving large-scale data retrieval. The hierarchical management strategy consists of the following three functional layers: 

\begin{itemize}[left=0pt, labelsep=0.5em, itemindent=0pt, listparindent=0pt]
  \item \textbf{Intra-tool Truncation and Metadata Logging.} The first layer operates at the individual tool level by implementing a rigid character-based constraint. When the raw output of a single tool exceeds a predefined threshold, the system executes an initial truncation to prevent a single retrieval action from exhausting the context budget. Crucially, the agent records the corresponding metadata during this process. This metadata serves as a structural anchor, allowing the system to track the origin and nature of the truncated content for more precise processing in subsequent layers.

  \item \textbf{Intra-round Adaptive Summarization.} The second layer focuses on the cumulative output within a single interaction round. This mechanism is triggered when the total length of all tool responses and their associated sub-queries reaches a specific heuristic threshold. In such cases, the system performs an intelligent synthesis of the round’s entire execution trace. If the total length significantly exceeds the threshold and contains previously truncated data, the summarization engine prioritizes the precise reconstruction of these truncated segments. This ensures that the agent retains a high-level semantic understanding of the round without being overwhelmed by raw, unformatted data.

  \item \textbf{Cross-round Retroactive Compression.} The third layer manages long-term history through a retroactive compression protocol. This process is initiated when the cumulative historical context reaches its capacity and the system identifies existing records of truncated tools. The mechanism backtracks through the interaction history to locate these truncated nodes and applies a secondary summarization. The original raw results are then replaced in-place with these compressed versions. This retroactive update significantly reduces the overall context footprint while preserving the critical logical links required for multi-step reasoning. To prevent the loss of information through repeat summarization, the system resets its truncation registry immediately after a cross-round update is completed. 
\end{itemize}

\subsection{Self-Evolving Semantic Memory}
We first propose self-evolving semantic memory named SES-Memory, for extracting transferable skills from execution trajectories. Unlike conventional memory systems constrained by binary success/failure evaluation, SES-Memory's process-centric skill extraction mimics human experiential learning, enabling continuous capability refinement. Specifically, traditional approaches extract full solutions only from successful tasks. SES-Memory retrieves high-value skill snippets from all agent execution trajectories. It does so regardless of whether the final task succeeds or fails. This mimics how human experts learn equally from all the past works because even the failed tasks may contain valuable insights in design or implementation. For instance, a data analysis task might fail due to incorrect final conclusions, but its intermediate steps could still have valuable information. SES-Memory can extract and generalize these local highlights as independent skills.

Through repeated attempts, agents automatically acquire rich prior knowledge from historical executions. SES-Memory is not limited to only history summary; In addition, they encompass diverse forms of reusable knowledge distilled from experience, including previously used code snippets, tools and libraries, technical insights, critical decision steps. Agents can further recombine elements from different the past executions to solve new tasks in alternative and more efficient ways.

Additionally, to ensure memory quality and scalability, SES-Memory incorporates a two-stage control mechanism. After retrieving the top‑k relevant memories, a similarity-based threshold filter is applied to prevent low-value or noisy memories from entering the context. Moreover, when a query retrieves multiple memories with consistently high similarity, the newly generated memory is not stored, thereby avoiding redundant entries and uncontrolled memory growth.

\subsection{Tool Augmentation}
To address the growing complexity and diversity of user interaction demands, Lemon Agent systematically enhances its tool set to provide agents with real-world problem-solving abilities. By equipping agents with improved perception, analysis, and retrieval tools, our system can effectively acquire, process, and synthesize information across different modalities and sources. This not only broadens the scope of tasks that Lemon Agent can handle, but also greatly improves its adaptability and reliability in real-world environments. 

\subsubsection{Intelligent Image Tool}
Current Vision-Language Models (VLMs) used for image understanding face inherent resolution limitations. On one hand, input images exceeding specific thresholds are typically resized before feature extraction and comprehension, which leads to a significant loss of fine-grained details. On the other hand, these models often exhibit limited perception of non-primary regions within an image. These two deficiencies collectively prevent VLMs from accurately analyzing target information in small localized areas when serving as visual perception tools. Notably, the spatial grounding and localization capabilities of modern VLMs have demonstrated continuous improvement. Leveraging this advantage, we design an intelligent image tool to enhance the comprehension of small-scale localized regions. The workflow begins with the VLM determining whether a localized analysis is required. If such refinement is deemed unnecessary, the model outputs the result directly. Conversely, if a localized view is required, the VLM generates diagonal relative coordinates for the target region, denoted as $r = [x_{min}, y_{min}, x_{max}, y_{max}]$, where the coordinates are normalized between $0$ and $1$. These relative values are then converted into absolute pixel coordinates $P = [X_{min}, Y_{min}, X_{max}, Y_{max}]$ based on the original image dimensions $(W, H)$ via the following transformation:
$$X_{i} = x_{i} \cdot W, \quad Y_{i} = y_{i} \cdot H$$
The original image is subsequently cropped according to these absolute coordinates, and the resulting high-resolution localized fragment will be re-input into the VLM to obtain the final refined result.

\subsubsection{Street View Agent}
To bridge the gap between high-level sub-tasks and low-level spatial navigation of google map, we introduce the Street View Agent, a specialized module designed to handle complex geospatial tasks within 3D environments. While browser-use agents often struggle with the high latency and redundant information of general web browsing, the Street View Agent serves as a dedicated interface that translates abstract navigational goals into precise API-driven actions. This mechanism ensures that the main agent remains focused on high-level strategy while the Street View Agent manages the intricacies of spatial perception and coordinate-based movement. The technical superiority of our Street View Agent is characterized by the following three core advantages: 

\begin{itemize}[left=0pt, labelsep=0.5em, itemindent=0pt, listparindent=0pt]
  \item \textbf{Panoramic Perception.} A primary challenge in spatial reasoning is the excessive number of interaction rounds required to explore an environment via standard interfaces. To address this, our agent implements a Panoramic Information Synthesis mechanism. Instead of sequential viewpoint adjustments, it retrieves geospatial metadata via the Google Maps API to programmatically stitch four orthogonal perspective images into a single panoramic representation. Our agent provides the LLM with a comprehensive environmental snapshot in a single round. 
  \item \textbf{Precise Coordinate-Based Navigation.} Unlike heuristic-based navigation in web browsers, which often relies on imprecise directional clicks, we implement Coordinate-Driven Spatial Control. For every movement, the agent explicitly calculates the target latitude and longitude based on the desired displacement. By directly interfacing with the API to fetch imagery at these precise coordinates, it achieves centimeter-level positioning. This granular control allows the agent to adaptively execute micro-movements for meticulous local searches.
  \item \textbf{Target-Centric Positioning.} We further integrate the newly proposed Image Tool to facilitate active visual grounding and spatial relocalization. When a candidate target is detected within the panoramic representation, the agent invokes the adaptive cropping mechanism to generate a high-resolution localized fragment of the scene. By deriving the precise bearing from the absolute coordinates calculated during the refinement process, the agent translates these pixel-level spatial data points into directional navigational vectors. This integration allows the agent to autonomously transition to the optimal vantage point where the target is most visible and least obstructed. The synergy between the refinement tool and the navigational interface ensures the collection of high-fidelity visual evidence required for successful task execution.
\end{itemize}

\subsubsection{Multi-Source Search Tool Suite}
To facilitate high-fidelity information grounding, Lemon Agent utilizes a diversified search suite comprising Google Search, Wikipedia historical revisions, and specialized web crawling capabilities. The core of this tool suite is a robust three-tier fallback protocol that prioritizes the Jina.ai Reader API for structured extraction, reverts to the Serper MCP service for broad-spectrum queries, and utilizes native HTTP requests as a final contingency. This hierarchical design is reinforced by an exponential backoff retry algorithm and an intelligent input sanitization layer that employs regular expressions to automatically rectify formatting errors or suboptimal query structures. Beyond raw retrieval, the system enhances the signal-to-noise ratio by dynamically filtering redundant information and providing the agent with granular diagnostic feedback based on specific error types. By translating failures into actionable suggestions for subsequent planning cycles, this closed-loop search solution ensures that Lemon Agent maintains a stable and coherent reasoning trajectory even when interfacing with unpredictable external data sources.

\subsection{Task Execution Procedure}
The Algorithm~\ref{alg:lemon_agent_pipeline} demonstrates the operational workflow of Lemon Agent through a structured sequence that integrates the architectural components described above. Lemon Agent starts the workflow by initializing the AgentCortex framework including the Agent Module and Execution Service. The system performs intent recognition to identify the global objective and recalls technical insights from the SES Memory. The system then loads specific skills and functional libraries based on these recalled insights before entering the reasoning loop.

Within the loop, Lemon Agent executes task routing and verification to determine the collaboration pattern through macro-level orchestration. The system deploys dynamic subagents to handle independent subtasks or assigns a single sub-worker for linear execution. The execution engine invokes MCP tools via streamable http protocols and selects between single tool calls or dynamic parallel tool calling. The system triggers specialized modules such as the intelligent image tool or the Street View Agent to process multimodal and geospatial data.

The system monitors the cumulative token count and applies the three layer context compression strategy when the predefined threshold is reached. This process includes intra tool truncation plus intra round summarization and cross round retroactive updates to maintain architectural stability. Lemon Agent synthesizes the final answer with a comprehensive overall analysis and a confidence score upon task completion. The system concludes the cycle by extracting high value skill snippets from the execution traces and storing these fragments back into the SES Memory for continuous experiential evolution.
\begin{algorithm}[tb]
    \caption{Lemon Agent Execution Pipeline}
    \label{alg:lemon_agent_pipeline}
    \begin{algorithmic}[1]  
        \REQUIRE User Task $T$
        \ENSURE Final Answer $A$ with confidence score $S_{\text{conf}}$
        
        \STATE \textbf{Using} AgentCortex.Framework: instantiate core services (Agents, MCP tools, Execution)
        
        \STATE \textbf{SES Memory Loading}: Retrieve top-$k$ relevant entries $\mathcal{M}_{\text{SES}}$ for $T$
        
        \STATE \textbf{Optional Reasoning}: \textit{If enabled}, perform reasoning on $T$ to better understand task intent
        
        \STATE \textbf{Main Loop (Task Routing \& Dynamic Subagents)}:
        \STATE \quad Initialize agent set $\mathcal{A} \leftarrow \emptyset$
        \STATE \quad \textit{While} task $T$ is not fully resolved:
        \STATE \quad \quad \textbf{Task Routing}: Route $T$ to appropriate subagents $\mathcal{A} = \{a_1, a_2, \ldots, a_n\}$
        \STATE \quad \quad \textit{For each} subagent $a_i \in \mathcal{A}$ \textbf{in parallel}:
        \STATE \quad \quad \quad Initialize context $C_i$ for subagent $a_i$
        \STATE \quad \quad \quad \textit{While} subagent $a_i$ not terminated:
        \STATE \quad \quad \quad \quad \textbf{Independent ReAct}: Let $a_i$ react to its assigned subtask
        \STATE \quad \quad \quad \quad \textbf{Dynamic Parallel Tool Calling}: 
        \STATE \quad \quad \quad \quad \quad Select tools $\mathcal{T}_i = \{t_1, t_2, \ldots, t_m\}$ for $a_i$
        \STATE \quad \quad \quad \quad \quad Execute all tools in $\mathcal{T}_i$ in parallel, collect results $\mathcal{R}_i$
        \STATE \quad \quad \quad \quad \textbf{Context Compression}:
        \STATE \quad \quad \quad \quad \quad 1. \textbf{Long reply truncated}: 
        \STATE \quad \quad \quad \quad \quad \quad \textit{For each} result $r_j \in \mathcal{R}_i$:
        \STATE \quad \quad \quad \quad \quad \quad \quad \textit{If} $|r_j| > \tau_{\text{max}}$: Truncate $r_j$ with marker
        \STATE \quad \quad \quad \quad \quad 2. \textbf{Intelligent tool summary}: 
        \STATE \quad \quad \quad \quad \quad \quad \textit{If} multiple tools and $\sum_{r_j \in \mathcal{R}_i} |r_j| > \tau_{\text{multi}}$:
        \STATE \quad \quad \quad \quad \quad \quad \quad Summarize $\mathcal{R}_i$ intelligently before feeding to $a_i$
        \STATE \quad \quad \quad \quad \quad 3. \textbf{Historical summary}: 
        \STATE \quad \quad \quad \quad \quad \quad \textit{If} $|C_i| > \tau_{\text{context}}$:
        \STATE \quad \quad \quad \quad \quad \quad \quad Compress historical context $C_i$ via summarization
        \STATE \quad \quad \quad \quad Update context $C_i$ with processed results
        \STATE \quad \quad \quad \textbf{Verification}: Subagent $a_i$ decides whether to verify its results
        \STATE \quad \quad \quad \textbf{Subagent Summary}: Subagent $a_i$ summary its results to reduce main agent's context length
        \STATE \quad \quad \quad Integrate subagent $a_i$ outputs
        
        \STATE \textbf{Result Aggregation}: Aggregate answers from all subagents to generate Final Answer $A$ with confidence score $S_{\text{conf}}$
        \STATE \textbf{SES Memory Writing}: Write execution context, task $T$, answer $A$, and score $S_{\text{conf}}$ to SES Memory
        
        \RETURN $A, S_{\text{conf}}$
    \end{algorithmic}
\end{algorithm}

\section{Experiments}

\subsection{Experimental Settings}
We evaluate Lemon Agent's general adaptability on the GAIA and xbench-DeepSearch Benchmark. \textbf{GAIA} represents a well-known benchmark designed to evaluate General AI Assistants on practical real-world problems requiring multiple core skills \citep{DBLP:conf/iclr/MialonF0LS24}. These questions remain conceptually simple for humans yet pose significant difficulties for the most advanced AI systems today. \textbf{xbench-DeepSearch} is part of xbench's AGI-Aligned series, evaluating tool usage capabilities in search and information retrieval scenarios \citep{xbench}. It focuses on the integration of planning plus multi step searching and logical synthesis to solve expert level queries. 

In our experiments we adopt GPT 5 as the core language model backbone with specific parameter configurations. For the main agent we set the reasoning level to high and the maximum number of interaction rounds to 10. For the sub-workers we also set the reasoning level to high while allowing a longer execution sequence of up to 20 rounds. The summary module utilizes a medium reasoning level to maintain a balance between computational efficiency and output quality. We configure the model temperature at 1.0 to ensure output diversity and limit the maximum output tokens to 128,000. Our empirical strategy utilizes a pass@3 approach where Lemon Agent attempts each task 3 times. The evaluation process supports multi process parallel execution with a default of 3 concurrent processes to ensure scalability.

\subsection{Benchmark Results}
\begin{figure}[tb]
    \centering
    \includegraphics[width=0.8\columnwidth]{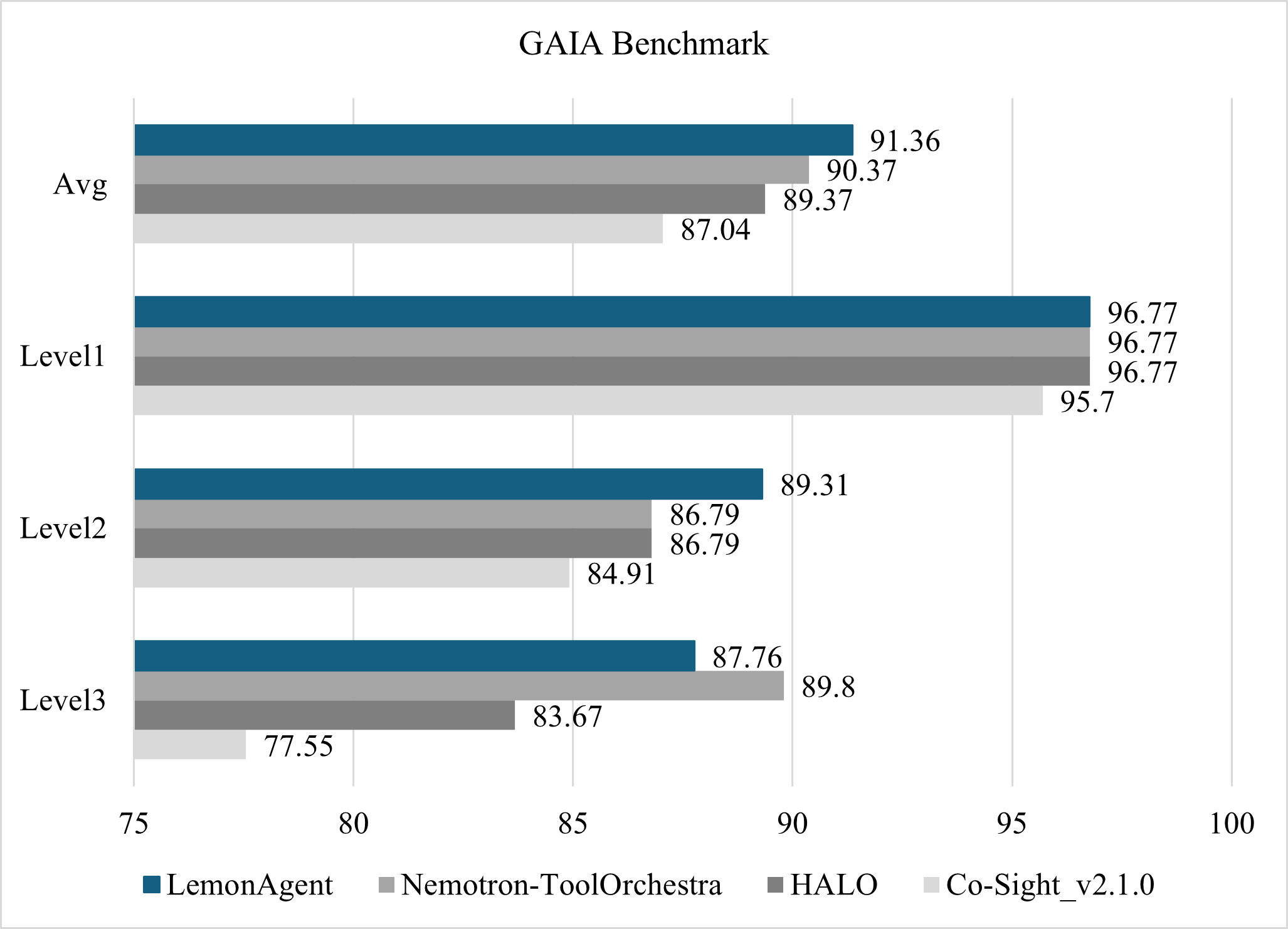}
    \caption{Comparative performance of Lemon Agent against other well-performed agents across all three difficulty levels of the GAIA benchmark.}
    \label{fig:gaia_results}
\end{figure}

As of February 6, 2026, Lemon Agent achieves 91.36\% overall accuracy on the GAIA benchmark across its three levels. Its per-level results are 96.77\% on Level 1, 89.31\% on Level 2, and 87.76\% on Level 3 tasks. Lemon Agent establishes state-of-the-art performance on both Level 1 and Level 2 among comparisons with other agents. These performance results are visualized in Figure~\ref{fig:gaia_results}, which illustrates substantial advances in multi step reasoning capabilities for complex extended task sequences.

Furthermore, Lemon Agent secures the top position on the xbench-DeepSearch leaderboard as of February 6, 2026. Table~\ref{tab:xbench_results} summarizes the comparative performance against other leading systems. It achieves an overall accuracy of 77+, surpassing the leading models including ChatGPT 5 Pro and SuperGrok Expert. This result shows the Lemon's superior ability to manage massive search spaces and synthesize high fidelity evidence from multi source information streams. 
\begin{table}[htbp]
    \centering
    \caption{Performance of Lemon Agent on the xbench-DeepSearch leaderboard.}
    \label{tab:xbench_results}
    \begin{tabular}{lccc}
        \toprule
        \textbf{Product} & \textbf{Organization} & \textbf{Accuracy} & \textbf{Evaluation Date} \\
        \midrule
        Minimax Agent & Minimax & 35+ & 2025-11-04 \\
        StepFun Research & StepFun & 35+ & 2025-08-28 \\
        Flowith & Flowith & 35+ & 2025-08-30 \\
        SuperGrok Expert & xAI & 40+ & 2025-08-28 \\
        ChatGPT-5-Pro & OpenAI & 75+ & 2025-08-28 \\
        \textbf{Lemon Agent} & \textbf{LR AILab of Lenovo CTO Org} & \textbf{77+} & \textbf{2026-02-04} \\
        \bottomrule
    \end{tabular}
\end{table}

\subsection{SES-Memory Results}

The SES-Memory module analyzes how AI agents complete tasks and extracts reusable skills from these experiences. It transforms specific code snippets into general patterns that can be applied to similar problems. For instance, detailed audio identification steps become a reusable template like "identify audio using Shazam." Common workflow sequences such as searching the web, crawling content, and recognizing text in PDFs are recognized as repeatable patterns. Our method also captures decision-making logic, creating rules like "use approach A when the task requires X, otherwise use approach B." These extracted skills are stored as structured natural language descriptions that remain easy for humans to understand while preserving actionable technical knowledge.

\begin{figure}[htbp]
    \centering
    \includegraphics[width=0.9\textwidth]{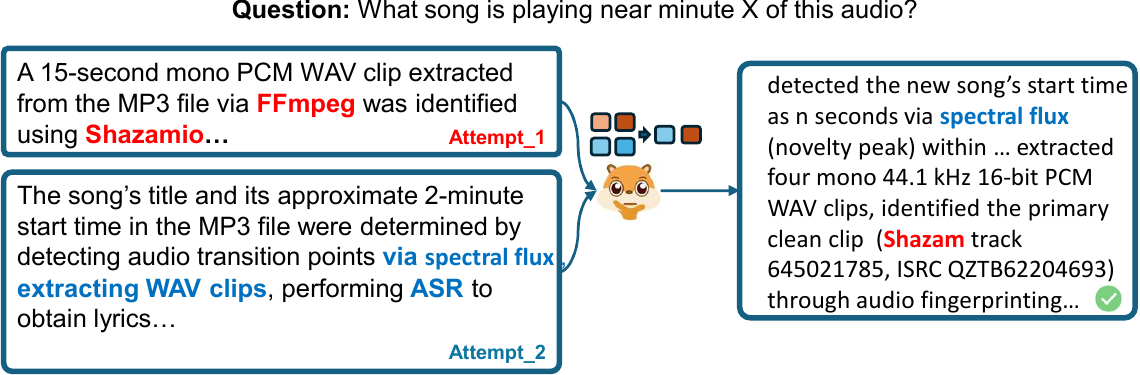}
    \caption{A case of completing a task by combining useful skills from multiple historical attempts via the SES-Memory module.}
    \label{fig:self-involved_memory}
\end{figure}

As shown in Figure~\ref{fig:self-involved_memory}, consider how this works when identifying music. During the first attempt, the agent used ffmpeg to extract music clips and shazamio to identify them. In the second attempt, it tried a different approach using spectral flux for audio extraction and ASR (Automatic Speech Recognition) for identification. For the third attempt, the agent leveraged insights stored in SES-Memory from both previous attempts, combining spectral flux for extraction with the Shazam library for identification to successfully complete the task. Importantly, SES-Memory doesn't require loading complete code libraries. Instead, it stores practical coding knowledge as simple prompts and instructions within its memory structure. This approach enables the system to improve performance on complex tasks without the overhead of managing extensive code repositories.

\subsection{Tool Augmentation Results}
\begin{figure}[htbp]
    \centering
    \def\subfigWidthA{0.520\textwidth}  
    \def\subfigWidthB{0.420\textwidth}   
    \def\figHeight{6cm}                  
    \def\figGap{0.03\textwidth}          
    
    \begin{subfigure}{\subfigWidthA}
        \centering
        \includegraphics[height=\figHeight, keepaspectratio]{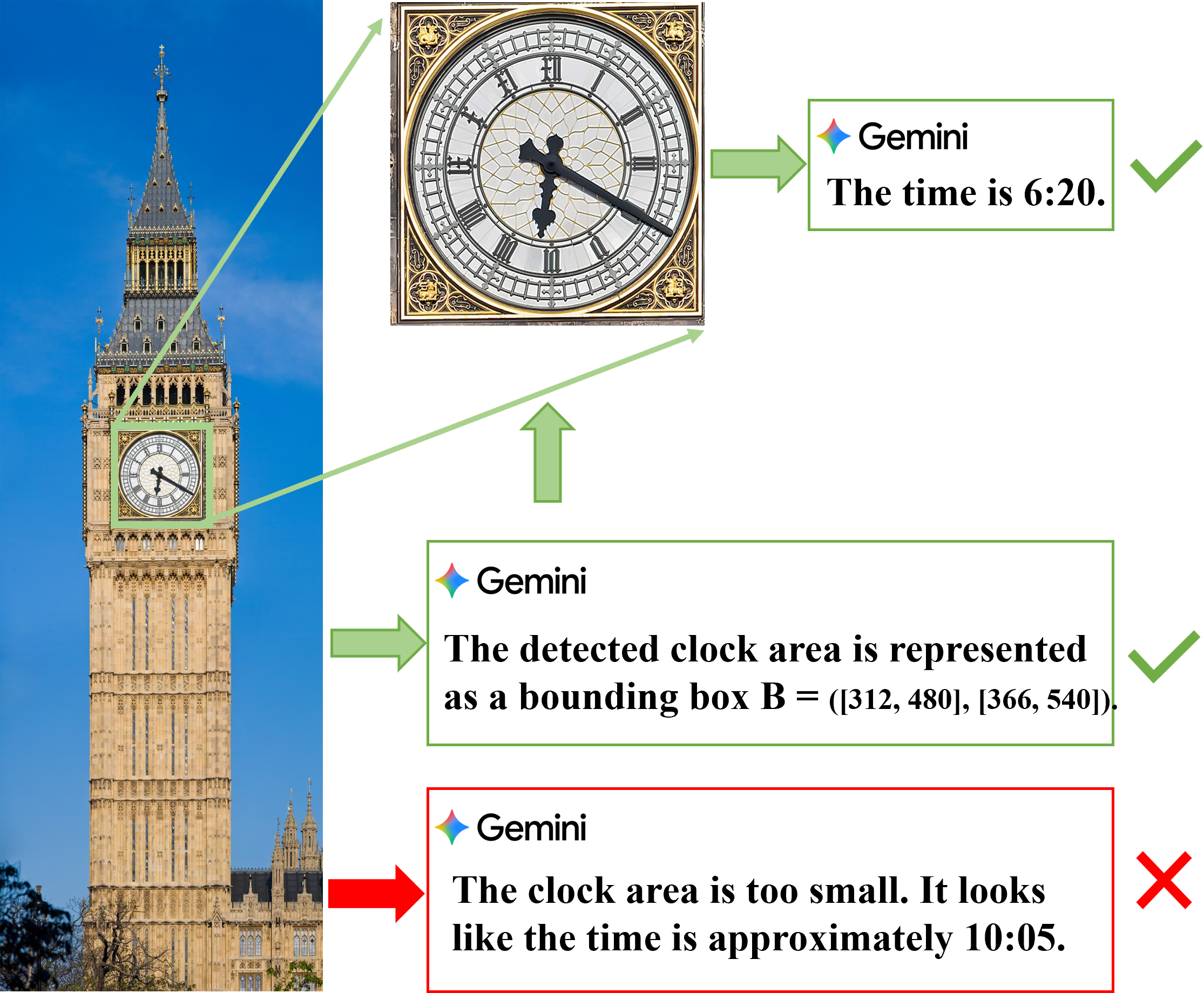}
        \caption{Intelligent Image Tool}
        \label{fig:exp-image_tool}
    \end{subfigure}
    \hspace{\figGap}
    \begin{subfigure}{\subfigWidthB}
        \centering
        \includegraphics[height=\figHeight, keepaspectratio]{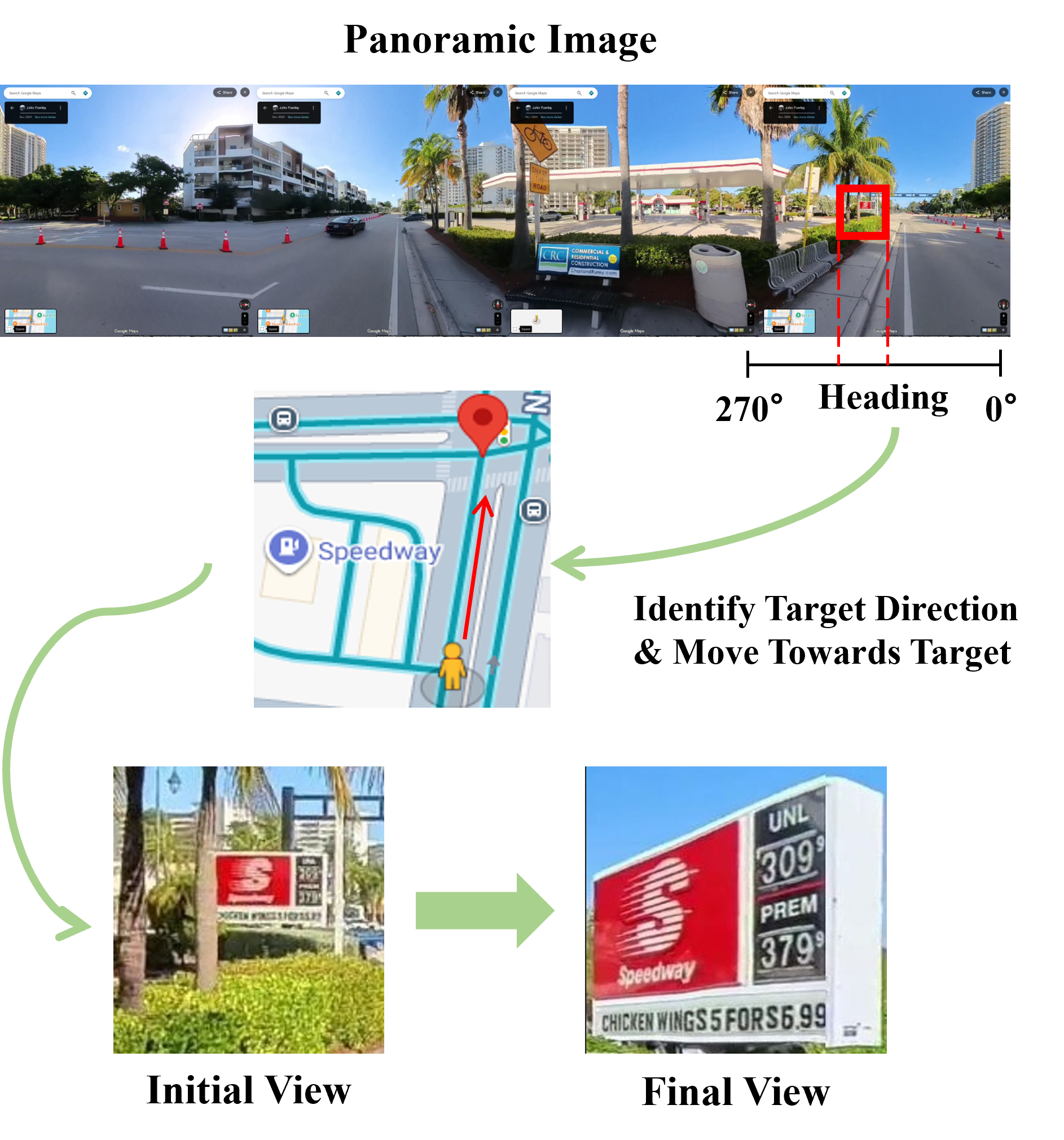}
        \caption{Street View Agent}
        \label{fig:exp-Street View Agent}
    \end{subfigure}
    \caption{GAIA cases illustrating the performance of the intelligent image tool and Street View Agent.}
\end{figure}

Our intelligent image tool demonstrates significant empirical improvements in fine grained visual perception. We show its effectiveness using a representative Level 2 task from the GAIA benchmark. This task requires calculating the temporal difference between two clocks. When we process original images directly using a standard vision language model such as Gemini 3 Pro, the results are frequently inaccurate in scenarios where the target object occupies a minimal proportion of the total image area. This failure stems from aggressive resolution compression where the model cannot distinguish between the hour and minute hands or completely misidentifies their orientations. Conversely, our adaptive tool dynamically isolates and crops the clock faces into high resolution localized fragments. This refinement allows the agent to extract precise visual evidence and reach the correct conclusion by bypassing the limitations of global image resizing. We illustrate this specific localized refinement procedure and its impact on reasoning accuracy in Figure~\ref{fig:exp-image_tool}.

Additionally, we show the performance of the Street View Agent through a complex geospatial reasoning task from the GAIA benchmark. Figure~\ref{fig:exp-Street View Agent} illustrates a Level 2 task that requires identifying fuel price information on a gas station signboard. Our Street View Agent begins the process by retrieving the panoramic image of the target location. To improve recognition precision, the agent first crops the panorama into candidate regions where the target may exist and performs an initial recognition to confirm the target's presence. In this example, once the gas station sign is detected, the agent calculates the required movement angle based on the relative coordinates of the crop within the panoramic image. Then it navigates along the road and precisely positions itself in the clearest viewpoint in front of the sign. Finally, the agent invokes the intelligent image tool to perform localized refinement on the retrieved sign. By cropping the sign into a high resolution fragment as evidenced by the transition from Initial view to Final view the agent accurately identifies the specific oil prices depicted on the board. This example successfully demonstrates the synergy between our spatial navigation modules and the adaptive visual refinement tool while confirming the generality of our Street View Agent across dynamic 3D environments.

\section{Limitation and Future Work}

While Lemon Agent achieves state of the art results across multiple benchmarks we recognize several constraints that guide our future research. Current tool interactions often require substantial data transfers between the agent and external environments which increases communication overhead and latency. To address this limitation we investigate the integration of sandboxed execution environments to enable agents to invoke modules through direct function calls. This design eliminates the necessity for repetitive uploads and downloads of large files such as images or structured data which significantly reduces context length. We also intend to explore more sophisticated multi-agent collaboration topologies to handle even more fragmented and orthogonal sub-goals. By performing tool operations locally within a sandbox and enhancing the precision of the SES Memory module we aim to further improve the efficiency and scalability of the overall system.

\section{Conclusion}

Lemon Agent has advanced deployment at industrial scale by integrating multi-agent collaboration with practical engineering considerations. We introduce a modular system that facilitates cost-effective, sustainable deployment while ensuring robust performance across varied operational environments. Our proposed system successfully coordinates hierarchical scheduling plus progressive context management and self evolving semantic memory within a unified Planner Executor Memory paradigm. Experimental results on the GAIA and xbench DeepSearch benchmarks verify the superior reasoning capabilities of Lemon Agent and its ability to manage massive search spaces. Moreover the successful deployment in the Lenovo Super Intelligent Agent proves the industrial grade reliability and commercial value of our approach. We provide the complete implementation as an open source project to foster community collaboration and support the advancement of autonomous agent research. This work represents a meaningful step toward balanced AI systems that prioritize both high fidelity capability and long term sustainability.


\clearpage
\bibliography{references}
\bibliographystyle{iclr2026_conference}

\end{document}